# TEMPERATURE DEPENDENCE OF THE COHERENCE IN POLARITON CONDENSATES

E. Rozas[1,2], M.D. Martín[1,2], C. Tejedor[2,3,4], L. Viña[1,2,4], G. Deligeorgis[5], Z. Hatzopoulos[5] and P.G. Savvidis[5,6,7]

[1]*Dept. Física de Materiales & Inst. N. Cabrera, Univ. Autónoma, Madrid 28049, Spain*
[2]*Instituto Nicolás. Cabrera, Univ. Autónoma, Madrid 28049, Spain*
[3]*Dept. Física Teórica de la Materia Condensada. Univ. Autónoma, Madrid 28049. Spain*
[4]*Inst. de Física de la Materia Condensada, Universidad Autónoma, Madrid 28049, Spain*
[5]*FORTH-IESL, P.O. Box 1385, 71110 Heraklion, Crete, Greece*
[6]*Dept. of Materials Science & Technology, University of Crete, 71003 Heraklion, Greece*
[7]*ITMO University, St. Petersburg 197101, Russia*

We present a time-resolved experimental study of the temperature effect on the coherence of traveling polariton condensates. The simultaneous detection of their emission both in real- and reciprocal-space allows us to fully monitor the condensates' dynamics. We obtain fringes in reciprocal-space as a result of the interference between polariton wavepackets (WPs) traveling with the same speed. The periodicity of these fringes is inversely proportional to the spatial distance between the interfering WPs. In a similar fashion, we obtain interference fringes in real-space when WPs traveling in opposite directions meet. The visibility of both real- and reciprocal-space interference fringes rapidly decreases with increasing temperature and vanishes. A theoretical description of the phase transition, considering the coexistence of condensed and non-condensed particles, for an out-of-equilibrium condensate such as ours is still missing. Yet a comparison with theories developed for atomic condensates allows us to infer a critical temperature for the BEC-like transition when the visibility goes to zero.

## I. INTRODUCTION

At low temperature, the optical properties of semiconductor crystals are dominated by exciton-polaritons, which are half-light, half-matter particles, resulting from the strong coupling between exciton and photon states [1]. Since their observation by Weisbuch *et al*. in semiconductor microcavities [2], polaritons in confined, low-dimensional structures have been profusely investigated. The strong light-matter coupling in the cavities gives rise to fascinating new effects that make polaritons appropriate candidates for nonlinear optical technologies [3]. Their excitonic part leads to strong polariton-polariton Coulomb interactions; furthermore, thanks to their photonic content, they can be easily created, by excitation with laser sources, and detected through the photon emission when polaritons annihilate. Their bosonic nature, together with their low effective mass and reduced density of states, facilitates their condensation in a macroscopic coherent phase that presents high similarities to Bose-Einstein condensates (BECs) [4]. Coherence is a key ingredient of condensates, and has been intensely studied both in atomic BECs [5], and in polariton condensates [6-23]. The coherence has been mostly investigated in real-space, by studying interference effects, either for static condensates [11, 24, 25] or for moving ones when they meet in real-space [10, 26]. Notwithstanding, a recent experimental study has focused on the study of coherence when two condensates, that move with the same speed, interfere in momentum (k)-space [27], circumventing the need of an encounter in real-space. Antón *et al*. showed the presence of interference fringes produced by the correlation between two components of such condensates that were spatially separated by 70 µm, demonstrating the existence of remote coherence between these condensates. Generally, the study of coherence has been performed at temperatures well below the critical temperature for condensation ($T_C$), to optimize the creation and stability of the condensates. To the best of our knowledge, only Bloch *et al*. [28] and Gati *et al*. [29, 30] have investigated the degree of spatial coherence as a function of T, in trapped atomic condensates. Analogous theoretical studies in the field of polariton condensates have been done [31]. Actually, the study of temperature dependence is arousing great interest in the field of polariton condensates. Lebedev *et al*. have analyzed the effect of finite temperature on the coupling of quasi-equilibrium exciton-polariton condensates in a Josephson junction: They describe a second order phase transition between classical (thermal) and quantum regimes characterized by a temperature parameter related to the polariton-polariton interaction length [32]. Moreover, another recent experiment by Ouellet-Plamondon *et al*. also gives the possibility to evaluate $T_C$ [33]. They reported on the dependence of polariton bistability with temperature, proposing that an increase in T leads to a significant incoherent population growth in the reservoir, which interacts with the polariton population. The study showed a collapse of the polariton hysteresis loop as well as a decrease of the transmitted intensity above the upward threshold of the hysteresis loops for a temperature of ∼ 22 K, revealing a strong loss of the coherent polariton population with increasing temperature. Here, we investigate the temperature dependence of the degree of coherence of polariton condensates in semiconductor microcavity ridges by a detailed analysis of interference patterns. Our measurements show a loss of coherence with increasing temperature both in real- and k-space.

## II. EXPERIMENTAL RESULTS AND DISCUSSION

The sample used in this work is a high-quality, Q-factor ∼ 16000, GaAs-based 1D-microcavity, surrounded by two Bragg mirrors. On its surface, a pattern has been sculpted throughout the sample, consisting in ridge structures with dimensions 20 μm x 300 μm. The sample displays a Rabi splitting of 9 meV and we perform our experiments in a region with approximately zero detuning. Further details about the sample can be found in Ref. [34]. It is kept in a variable-temperature cryostat under high vacuum conditions. The temperature dependence of the degree of coherence is studied from 10 K up to a temperature of 42.5 K.

To study the coherence of the condensates, we excite the sample with 2 ps-long light pulses from a Ti:$Al_2O_3$ laser source. The laser beam is divided into two, which are focused on the sample through a microscope objective. Both beams are precisely controlled to have the same power density; they arrive simultaneously, impinging at the same angle, and are separated by d = 70(1) μm on the sample surface. We adjust the optical excitation so that particles are initially created with $k_{//}$~0. The angle of detection ($\theta$) is directly related to the parallel component of the polariton momentum, $k_{//}$, by $k_{//}= n\frac{2\pi}{\lambda}\sin\theta$.

Low temperature (10 K) continuous wave photoluminescence (PL) measurements (not shown) reveal that the emission spectrum of the sample is composed of a broad band between 1.5480 eV and 1.5420 eV, originating from the excitonic recombination [35] together with several narrow lower polariton branches (LPBs), between 1.5420 eV and 1.5398 eV, arising from the confinement along the narrow dimension of the ridge. Under resonant excitation conditions the phase of the polariton condensates would be inherited from that of the laser pulses, therefore to study the genuine condensates' coherence we excite quasi-resonantly, at 1.5459 eV. We find that the threshold for polariton propagation out of the excitation spot is 3.7 kW/cm$^2$ and we perform our experiments with an excitation power of 10 kW/cm$^2$. The blueshift obtained under these conditions is approximately 0.6 meV. More details of the excitation conditions are reported in Ref [27] and Fig. 1 therein. The time-resolved PL is measured at the energy of traveling polaritons (1.5404 eV with a resolution of 0.45 meV), using a streak camera with an overall time-resolution of ∼ 10 ps, averaging over millions of laser pulses. We obtain emission maps vs position and time focusing the emission directly onto the entrance slit of a spectrometer connected to the streak camera. We are also able to record the PL vs momentum and time, with the aid of an additional lens in our setup, collecting the light distribution in the Fourier plane of the microscope objective [27].

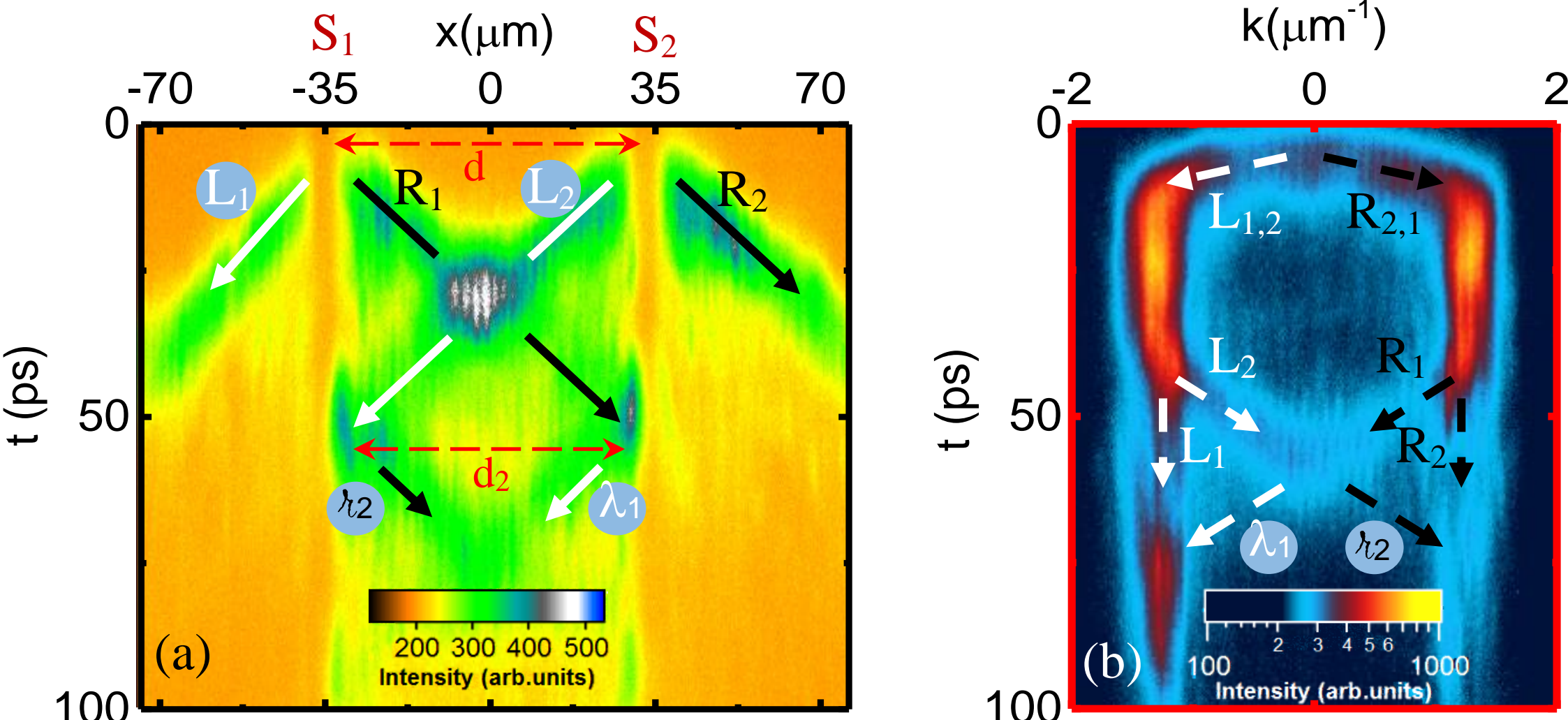

Figure 1. (a) PL emission along the ridge in real-space (x) as a function of time for T = 14 K. $S_{1,2}$ mark the positions where the two laser beams impinge on the sample. $L_i$ ($R_i$) denote the WPs moving to the left (right), with the subscript, i, referring to the excitation beam. The red dashed arrow at the top of the image represents the spatial separation (d) between both beams at t = 0 while the arrow at ~ 55 ps marks the separation ($d_2$) between WPs $R_1$ and $L_2$ when they arrive at the excitonic reservoirs. The intensity is in a linear false-color scale. (b) The corresponding emission in momentum-space (k) as a function of time. The color of the arrows refers to the WPs described in (a). In this case, the intensity is in a logarithmic false-color scale. Both PL emissions have been measured with a power density of 6 kW/cm$^2$.

For excitation densities above threshold for polariton condensation, and after energy relaxation, polaritons with k ~ 0 evolve towards two states with momenta ±k. As a result, the condensates propagate on both sides, away from the excitation spots [36, 37]. This dynamics under the two-beam excitation conditions used in our experiments is compiled in Fig. 1, which depicts the emission maps both in real- and momentum-space. In real-space, [Fig. 1(a)], two polariton wave packets (WPs) arise from each excitation spot ($S_1$ @-35 μm and $S_2$ @+35 μm), which move in opposite directions. We label these four WPs as $L_i/R_i$, as they move away from the excitation area $S_i$ (i=1, 2) towards the left/right respectively. The WPs move with a constant speed of ~ 1.5 μm/ps and their emission can be monitored up to ~ 130 ps. Dim fringes can be seen along the path of individual polariton wavepackets. They originate from the interference of the traveling polaritons with those backscattered by the disorder present in the sample [38]. We can single out these fringes from the ones occurring when two condensates meet since their periodicity is slightly larger due to the fact that backscattered polaritons travel with a wavevector slightly shorter than that of the forth propagating polaritons.

The outermost WPs, $L_1$ and $R_2$, exit the detection window after 45 ps and therefore their PL cannot be measured from there on. The other two WPs, $R_1$ and $L_2$, moving towards each other, meet in real-space (x ~ 0) at ~ 35 ps and interfere with each other, as evidenced by the interference fringes. At later times, $R_1$ and $L_2$, continue traveling towards $S_2$ and $S_1$, reaching the vicinity of these positions (x~±35 μm) at ~ 55 ps. There, the potential walls created by the excitonic reservoirs [36,37] act as barriers that the WPs cannot overcome, decelerating and eventually stopping and reversing their trajectories, so that $R_1$ ($L_2$) becomes $\lambda_1$ ($r_2$). The separation between $L_2$ and $R_1$ at this time, $d_2$, is slightly smaller than d because the WPs are not able to reach the maxima of the barriers. Figure 1(b) shows the corresponding time evolution of the WPs in momentum-space. Initially, the condensates accelerate from rest (k = 0) and they move left (right), reaching k = - 1.3 μm$^{-1}$ (k = +1.3 μm$^{-1}$) in a few picoseconds. The WPs travel with this wave vector up to ~ 40 ps. The left (right) trace reflects the movement of $L_1$ and $L_2$ ($R_1$ and $R_2$), respectively. From there on $L_2$ and $R_1$ start to decelerate, due to the presence of the excitonic reservoir potentials. Consequently, when the polaritons reach the vicinity of $S_1$ and $S_2$ (t ~ 55 ps) and stop, the traces merge at k ~ 0. Later on, the WPs reverse their movement direction [$L_2$ ($R_1$) becomes $r_2$ ($\lambda_1$)] and accelerate again until they attain the same momentum of the outermost WPs, $k = \pm 1.3$ μm$^{-1}$. We observe interference fringes during the first ~ 40 ps, both at negative and positive values of k, and later on, at ~ 55 ps and k~0. The former ones arise from the fact that $L_1$ and $L_2$ ($R_1$ and $R_2$) increase their speed with the same acceleration, therefore traveling with the same momentum until they reach - 1.3 μm$^{-1}$ (+1.3 μm$^{-1}$), fulfilling the requirements for interference in k-space. To observe these fringes, it is crucial to assure the same conditions for each excitation spot, creating equal excitonic populations that lead to equal kinetics and consequently equal momenta (i.e., overlapping of the wave vectors in k-space) of the WPs traveling in the same direction. It also requires keeping a constant distance between the beams in real-space, since it determines the period of the interference patterns, as will be discussed below. The second set of fringes (at t~55ps) appears when $R_1$ and $L_2$ both have very low speed in the environs of the excitonic reservoirs ($S_{2,1}$). From then on, the polariton finite lifetime starts to hinder the detection of the condensates' emission. However, interferences are still observable for those polaritons moving to the left when the outermost WP $L_1$ merges with $\lambda_1$.

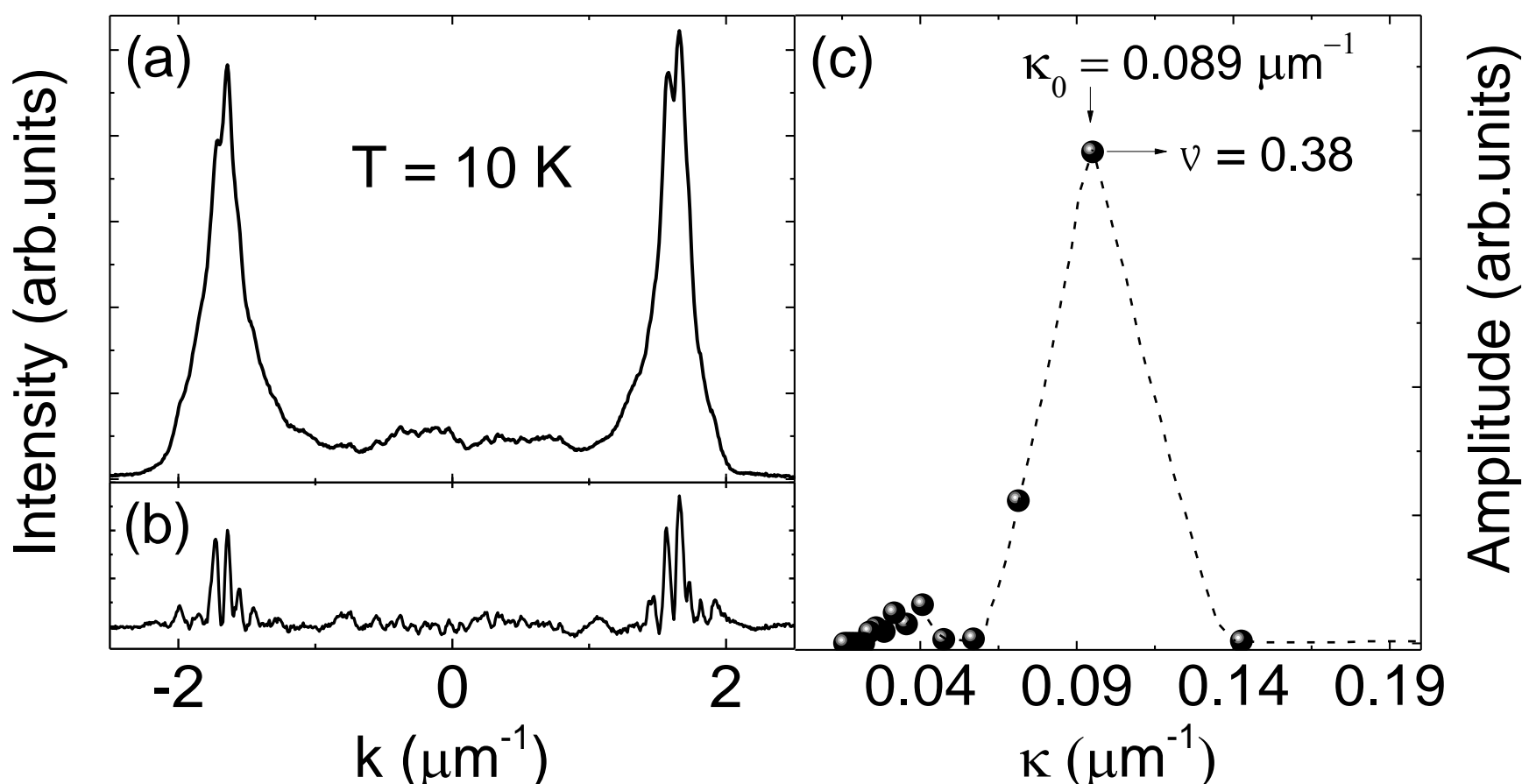

Figure 2. (a) PL emission in momentum-space time-integrated for $t_1$ at a temperature of 10 K. (b) Interferogram profile after a baseline subtraction of the trace shown in (a). (c) Amplitude of the different contributions to the interferogram obtained from a Fourier analysis of the trace depicted in (b): the value of the main period of the interference fringes is marked as $\kappa_0$; the corresponding visibility as $\nu$. The dashed line shows a Bézier interpolation of the points. A 10 kW/cm$^2$ excitation laser was used in the measurements.

Now, we study the temperature dependence of the mutual coherence between different WPs. Since our system is far from equilibrium it does not have a well-defined temperature, yet we can have some insight on the thermal robustness of the polariton condensate coherence by studying the evolution of the interference fringes' visibility with the lattice temperature. We will focus our attention on three time-windows of particular interest in which, during the whole time-spans, the amplitude of the interference fringes remains constant (within experimental accuracy). In momentum-space, we evaluate the interference obtained at two different time intervals; at $t_1$, from 13 ps to 39 ps, when the four WPs acquire a constant maximum momentum, and at the interval $t_3$, from 42 ps to 57 ps, where $R_1$ and $L_2$ have decreased their speed and reduced their momentum down to k ~ 0. We also investigate the interference pattern in real-space, resulting from the meeting of the WPs $R_1$ and $L_2$ around x ~ 0 at the interval $t_2$, from 22 ps to 37 ps.

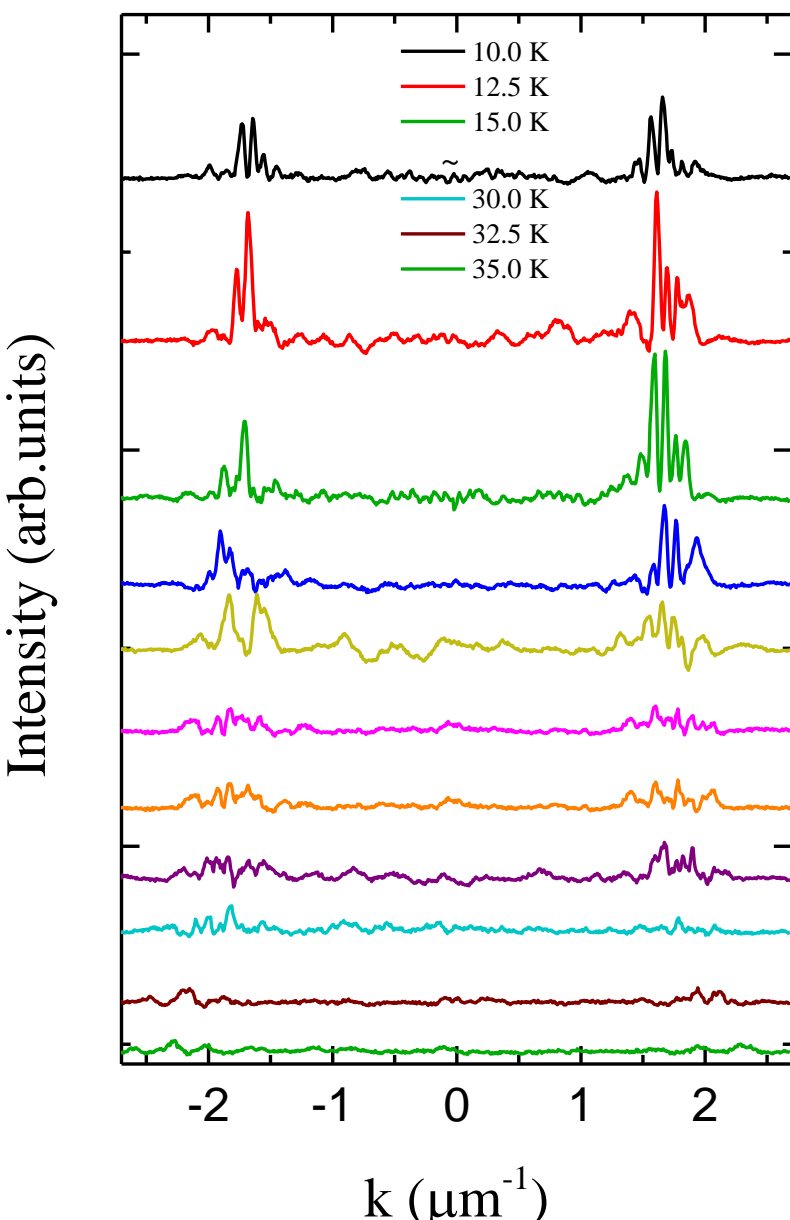

Figure 3. Profile of the emission spectra for $t_1$ in momentum-space varying the temperature from 10 K (top) to 35 K (bottom) in steps of 2.5 K (only the first and last three temperatures are labeled).

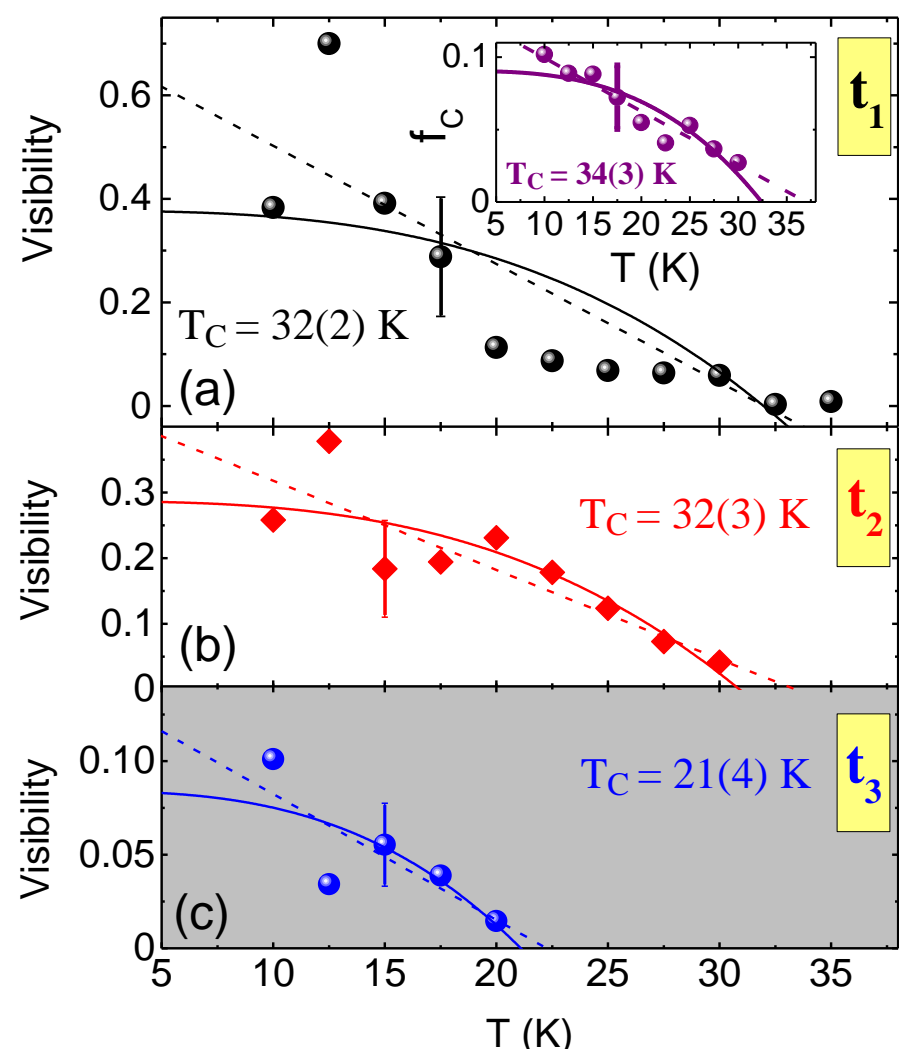

Figure 4. Temperature dependence of the visibility of the interference fringes: in momentum-space for the interval $t_1$ (black), in the vicinity of k = 0 at $t_3$ (blue) and in real-space at the first crossing of the wavepackets, $R_1$ and $L_2$, $t_2$ (red). The inset shows the fraction of condensed to uncondensed polariton populations ($f_C$) for increasing temperatures during the time interval $t_1$. The lines represent fits of the data with $v(T)=v_0[1-(T/T_C)^\beta]$, for β=1 (dotted line) and β=3 (solid line).

To analyze the data and obtain the period and amplitude of the interference fringes, we perform a detailed Fourier analysis of these patterns. The observed emission originates from both condensed and thermal, non-condensed, polaritons. The contribution of the latter is apparent in the data as a background on top of which the interference fringes are observed. In order to obtain a cleaner interferogram, we subtract a baseline to the patterns. As an example, Fig. 2(a) depicts the profile of the emission, integrated in the time interval $t_1$, in the full range of momenta, at T = 10 K, revealing clear oscillations related to the interference fringes. Figure. 2(b) displays the same profile after the baseline subtraction. Its Fourier analysis, depicted on Fig. 2(c) and obtained selecting the wavevector range $1.2 < |k| < 2.1$ μm$^{-1}$, obtains the amplitudes of the different periods present in the interference pattern, $\kappa \equiv \Delta k$, and reveals a peak corresponding to the predominant period in momentum-space, $\kappa_0$. This period is related to the distance (d) between both excitation spots, $S_1$ and $S_2$, by $\kappa_0 = 2\pi/d$ [27]. The visibility of the fringes shown in Fig. 2(b), is also computed as $v = \frac{1}{n}\sum_{i=1}^{n}\left(\frac{I_{max}-I_{min}}{I_{max}+I_{min}}\right)_i$, where $I_{max}(I_{min})$ is the maximum (minimum) intensity of the interference oscillation $\#i$ of the $n$ oscillations observed in the interferogram. This value is used to scale the amplitude given by the Fourier analysis for each time interval, as depicted in Fig. 2(c).

We perform a similar analysis of the interferograms obtained for different temperatures. The profiles of the baseline-subtracted emission intensity vs k for $t_1$, measured at different temperatures, in steps of 2.5 K, are compiled in Fig. 3. The interference fringes are directly observed in the range of maximum momentum of the PL emission. Above ~ 32.5 K, the lifetime of the polaritons is drastically reduced, hindering their propagation, as demonstrated by the strong reduction of the PL at large |k| values. This also results in the disappearance of the interference fringes. The analysis yields a temperature-independent period

of the interferences $\kappa_0 = 0.089(6)$ μm$^{-1}$, indicating a distance in real-space between the moving condensates of 71(5) μm. The T-independence of $\kappa_0$ is expected from the experimental fact that the separation between the two excitation laser-beams is kept constant at d = 70(1) μm for all temperatures. Figure 4 displays the visibility of the fringes ($\upsilon$) vs T: $\upsilon$ decays indicating a loss of coherence with increasing temperature. We interpret this reduction of the visibility as a signature of a BEC-like transition and obtain the characteristic critical temperature, $T_C^{t_1} = 32\ (2)$ K, from its value when $\upsilon$ vanishes [see Fig. 4(a)]. To further support our findings, we have also obtained the fraction of condensed to un-condensed polariton populations ($f_C$) and evaluated its temperature dependence. We have computed $f_C$ as the ratio between the area underneath the interference fringes and the baseline and that enclosed by the baseline. As can be observed in the inset of Fig. 4(a), we obtain a similar $T_C^{t_1}$ within error bars.

We find a similar visibility reduction when we turn our attention to the interference pattern obtained at $t_3$, when $R_1$ and $L_2$ meet in momentum-space at k ~ 0, resulting in a second set of fringes. In this case, the Fourier analysis gives a period of $\kappa_0 = 0.106(2)$ μm$^{-1}$ corresponding to a distance of 59(1) μm. This can be directly checked looking at the real-space measurements that reveal a distance between WPs $R_1$ and $L_2$ of $d_2 = 60(1)$ μm [Fig. 1(a)]. For this time interval, the signal to noise ratio limits the temperature range in which a reliable Fourier analysis can be performed, so only temperatures below 20 K have been considered. The temperature dependence of the visibility, reveals a similar decay to that obtained for $t_1$. However, the visibility decay with temperature is much faster than for $t_1$, obtaining a critical temperature of $T_C^{t_3} = 21\ (4)$ K [see Fig. 4(c)]. Furthermore, for a given temperature, the value of the visibility at $t_3$ is significantly smaller than that obtained at $t_1$, revealing that the coherence gradually decreases with time as the polaritons travel along the sample. The robustness of coherence with temperature is reduced for this time interval since the WPs are in close proximity to the excitonic reservoirs, which contribute to a faster decoherence through exciton-polariton scattering [39-41].

For the sake of completeness, we have also analyzed the interference patterns observed in real-space at $t_2$, when $R_1$ and $L_2$ meet at x ~ 0 [see Fig. 1(a)]. Now, the period of the interference fringes, ξ, is related to the distance in k between WPs $R_1$ and $L_2$, given by $\xi = 2\pi/\left|\vec{k}_{R_1} - \vec{k}_{L_2}\right|$ [27]. From our Fourier analysis, we obtain ξ= 1.8(1) μm; i.e. $\left|\vec{k}_{R_1} - \vec{k}_{L_2}\right| = 3.4(2)$ μm$^{-1}$; and taking into account that $\left|\vec{k}_{R_1}\right| = \left|\vec{k}_{L_2}\right|$ for t < 42 ps, this yields $\left|\vec{k}_{R_1}\right| = \left|\vec{k}_{L_2}\right| = 1.7(1)$ μm$^{-1}$. This value is in very good agreement with the maximum k value $\left|\vec{k}_{R_1}\right| = \left|\vec{k}_{L_2}\right| = 1.70\ (2)$ μm$^{-1}$ directly measured [see Fig 1(b)]. In this case, the analysis can be performed just up to 30 K due to lifetime constraints. Red points in Fig. 4(b) compile the temperature dependence of the visibility in real-space: The general behavior of the coherence with increasing temperature is similar to that observed in momentum-space, as expected from the connection between real- and k-space. From the fits of these data, obtained at $t_2$, we found $T_C^{t_2} = 32(3)$ K.

Very few studies of the temperature dependence of condensates' coherence can be found in the literature. Two situations have been theoretically considered, to the best of our knowledge, for atomic condensates: i) a refined description of a 3D cold atom gas confined in a cigar like trap [42] and ii) a mean field approach for a purely 2D atom gas [43]. In both cases, the temperature dependence of the fraction of condensed atoms takes the general form:

$$n_C(T) = n_0\left[1 - \left(\frac{T}{T_C}\right)^{\beta}\right]$$

where $T_C$ is a transition temperature for condensation. In the former approach β is close to 3 while in the latter β ~ 1. Lacking a detailed theory for non-equilibrium polariton condensates, we test the validity of both atomic theories comparing the behavior of the temperature dependence of the condensed fraction for both models with our results for the visibility of the non-equilibrium polariton condensates. As shown in Fig. 4 both models fit reasonably well our experimental results. Therefore, it is difficult to discriminate between these two approaches and to conclude which model describes more appropriately the temperature dependence of the coherence. It should also be mentioned that in the theoretical models the condensate fraction (and therefore the visibility) is 1 at zero temperature, while our experiments obtain a low-temperature value of the visibility well below 1. This originates from the omnipresence of thermal, non-condensed polaritons [44] and from the non-equilibrium nature of the polariton condensates, implying that, even at extremely low temperature, there are significant effects arising from out-of-equilibrium noise. In any case, independently of the model used to account for the temperature dependence of the coherence, the experimental data directly yield $T_C$, with an accuracy of ± 3 K, from the temperature at which the visibility dies out.

## III. CONCLUSIONS

In conclusion, we have investigated the temperature dependence of the mutual coherence of polariton condensates created separately in real-space in semiconductor microcavity-ridges. At low temperature, a conspicuous phase correlation is observed during the full propagation of the WPs, revealed by the presence of interference fringes in the whole range of the momentum emission. The temperature dependence of the coherence has been evaluated, through the visibility of the interference patterns, at three different time-delays after the condensate formation. A similar decrease of the coherence with increasing temperature has been found, both in real- and momentum-space that is correlated with the BEC-like transition. A comparison with two theoretical models developed for atomic condensates does not provide a clear conclusion about which model is more appropriate. Moreover, significant differences can be expected when comparing theories developed for equilibrium condensates with polariton condensates, which demands a more adequate model that takes into account the non-equilibrium nature of the polariton condensates.

## ACKNOWLEDGMENTS

E.R. acknowledges financial support from a Spanish FPI scholarship No. BES-2015-074708. This work was partially supported by the Spanish MINECO grants No. MAT2014-53119-C2-1-R and No. MAT2017-83722-R. P.G.S. acknowledges support from ITMO Fellowship Program and megaGrant No. 14.Y26.31.0015 of the Ministry of Education and Science of Russian Federation.